\documentclass[11pt,draftclsnofoot,peerreviewca,onecolumn,a4]{IEEEtran}
\usepackage{amsmath, mathrsfs}
\usepackage{mathtools}
\usepackage{bm}
\usepackage{bbm}
\usepackage{stmaryrd}
\usepackage[dvipsnames]{xcolor}
\makeatletter
\def\maketag@@@#1{\hbox{\m@th\normalfont\normalsize#1}}
\makeatother

\newcommand{\subparagraph}{}
\usepackage[compact]{titlesec}
\titlespacing*{\section}{0pt}{1\baselineskip}{0.9\baselineskip}

\allowdisplaybreaks

\usepackage{tikz}
\usepackage{pgfplots}
\pgfplotsset{compat=newest}
\usetikzlibrary{patterns}
\usetikzlibrary{positioning}
\usetikzlibrary{datavisualization}
\usetikzlibrary{datavisualization.formats.functions}
\usetikzlibrary{backgrounds}
\usetikzlibrary{shapes,snakes}
\usepackage{textcomp}
\usetikzlibrary{positioning}

\usepgfplotslibrary{external} 
\usepackage{cite}
\usepackage{amsfonts}
\usepackage{amssymb}
\usepackage{color}
\usepackage{multirow}
\usepackage{graphicx}
 \usepackage{relsize}
\usepackage{multirow}
\usepackage{multicol}
\usepackage{caption}
\usepackage{subcaption}
\usepackage[numbers,sort&compress]{natbib}
\usepackage{balance}
\usepackage{arydshln}

\usepackage{pgfplots} 
\usepackage{pgfgantt}
\usepackage{pdflscape}
\pgfplotsset{compat=newest} 
\pgfplotsset{plot coordinates/math parser=false}
\newlength\fwidth

\def\mindex#1{\index{#1}}

\def\sq{\hbox{\rlap{$\sqcap$}$\sqcup$}}
\def\qed{\ifmmode\sq\else{\unskip\nobreak\hfil
\penalty50\hskip1em\null\nobreak\hfil\sq
\parfillskip=0pt\finalhyphendemerits=0\endgraf}\fi\medskip}

\long\def\defbox#1{\framebox[.9\hsize][c]{\parbox{.85\hsize}{%
\parindent=0pt
\baselineskip=12pt plus .1pt      
\parskip=6pt plus 1.5pt minus 1pt 
 #1}}}

\long\def\beginbox#1\endbox{\subsection*{}%
\hbox{\hspace{.05\hsize}\defbox{\medskip#1\bigskip}}%
\subsection*{}}

\def\endbox{}

\newsavebox{\junk}
\savebox{\junk}[1.6mm]{\hbox{$|\!|\!|$}}

\def\argmin{\mathop{\rm arg\, min}}

\def\bC{{\mathbb C}}

\def\bE{{\mathbb E}}

\def\bfB{{\bf B}}

\def\bfF{{\bf F}}
\def\bfG{{\bf G}}
\def\bfH{{\bf H}}
\def\bfI{{\bf I}}

\def\bfT{{\bf T}}

\def\bfX{{\bf X}}

\def\bfa{{\bf a}}

\def\bfh{{\bf h}}

\def\bfn{{\bf n}}

\def\bfr{{\bf r}}
\def\bfs{{\bf s}}

\def\bfx{{\bf x}}
\def\bfy{{\bf y}}

\def\sfH{{\sf H}}

\def\bfmath#1{{\mathchoice{\mbox{\boldmath$#1$}}%
{\mbox{\boldmath$#1$}}%
{\mbox{\boldmath$\scriptstyle#1$}}%
{\mbox{\boldmath$\scriptscriptstyle#1$}}}}

\def\bfmY{\bfmath{Y}}

\def\bfmhhaY{\bfmath{\hhaY}} 
\def\bfmhhaY{\hbox to 0pt{$\widehat{\bfmY}$\hss}\widehat{\phantom{\raise 1.25pt\hbox{$\bfmY$}}}}

\def\til={{\widetilde =}}

\def\clC{{\cal C}}

\def\clN{{\cal N}}

 \def\FRAC#1#2#3{\genfrac{}{}{}{#1}{#2}{#3}}

\def\ddtp{{\mathchoice{\FRAC{1}{d^{\hbox to 2pt{\rm\tiny +\hss}}}{dt}}%
{\FRAC{1}{d^{\hbox to 2pt{\rm\tiny +\hss}}}{dt}}%
{\FRAC{3}{d^{\hbox to 2pt{\rm\tiny +\hss}}}{dt}}%
{\FRAC{3}{d^{\hbox to 2pt{\rm\tiny +\hss}}}{dt}}}}

\def\average#1,#2,{{1\over #2} \sum_{#1}^{#2}}

\def\eye(#1){{\bf(#1)}\quad}

\def\eq#1/{(\ref{e:#1})}

\newcommand{\beqn}[1]{\notes{#1}%
\begin{eqnarray} \elabel{#1}}

\newcommand{\eeqn}{\end{eqnarray} }

\newcommand{\beq}[1]{\notes{#1}%
\begin{equation}\elabel{#1}}

\newcommand{\eeq}{\end{equation}}

\def\bdes{\begin{description}}
\def\edes{\end{description}}

\newcounter{rmnum}

\newcounter{anum}

{\end{list}}

\def\ass(#1:#2){(#1\ref{#1:#2})}

\def\ritem#1{
\item[{\sf \ass(\current_model:#1)}]
}

\newenvironment{recall-ass}[1]{%
\begin{description}
\def\current_model{#1}}{
\end{description}
}

\long\def\comment#1{}

\newfont{\bbb}{msbm10 scaled 700}

\newfont{\bb}{msbm10 scaled 1100}
\newcommand{\CC}{\mbox{\bb C}}

\newcommand{\EE}{\mbox{\bb E}}

\newcommand{\av}{{\bf a}}

\newcommand{\hv}{{\bf h}}

\newcommand{\nv}{{\bf n}}

\newcommand{\yv}{{\bf y}}

\newcommand{\Bm}{{\bf B}}

\newcommand{\Fm}{{\bf F}}
\newcommand{\Gm}{{\bf G}}

\newcommand{\Id}{{\bf I}}

\newcommand{\Nm}{{\bf N}}

\newcommand{\Qm}{{\bf Q}}

\newcommand{\Xm}{{\bf X}}
\newcommand{\Ym}{{\bf Y}}

\newcommand{\Ic}{{\cal I}}

\newcommand{\Kc}{{\cal K}}

\newcommand{\Oc}{{\cal O}}

\newcommand{\Sc}{{\cal S}}

\newcommand{\Xc}{{\cal X}}

\renewcommand{\arg}{{\hbox{arg}}}

\newcommand{\SINR}{{\sf SINR}}
\newcommand{\SNR}{{\sf SNR}}

\newcommand{\transp}{{\sf T}}

\newcommand{\bphi}{\boldsymbol{\phi}}

\usepackage{algorithm}
\usepackage{algpseudocode}
\usepackage{pifont}
\usepackage{varwidth}

\graphicspath{{./Figures/}}

\newcommand{\xiul}{{\mathcal{I}}_{{\rm ul},i}}
\newcommand{\xidl}{{\mathcal{I}}_{{\rm dl},i}}
\newcommand{\Xgam}{\mathcal{X}_{\gamma}}
\newcommand{\Xgamhat}{\hat{\mathcal{X}}_{\gamma}}

\newcommand{\aulvec}{\bfa_{\rm ul}(\theta)}
\newcommand{\aulveccheck}{\check{\av}_{\rm ul}(\theta)}
\newcommand{\adlveccheck}{\check{\av}_{\rm dl}(\theta)}

\newcommand{\adlvec}{\bfa_{\rm dl}(\theta)}
\newcommand{\Pbf}{P_{\small \text{\sf prob}}}

  \newcommand{\Bcup}{\mathlarger{\mathlarger{\cup}}}

\newcommand{\lambdaul}{\lambda_{{\rm ul}}}
\newcommand{\lambdadl}{\lambda_{{\rm dl}}}
\newcommand{\suppul}{{\cal S}_{\rm ul}}
\newcommand{\suppulhat}{\hat{{\cal S}}_{\rm ul}}

\newcommand{\suppdl}{{\cal S}_{\rm dl}}
\newcommand{\suppdlhat}{\hat{{\cal S}}_{\rm dl}}

\def\herm{{\sfH}}

\def\cg{{\clC\clN}}

\newtheorem{obs_env}{Observation}

\begin{document}

\title{Efficient Downlink Channel Probing and Uplink Feedback in FDD Massive MIMO Systems}
\author{Mahdi Barzegar Khalilsarai, \IEEEmembership{Member, IEEE,} Saeid Haghighatshoar,  \IEEEmembership{Member, IEEE,} \\Giuseppe Caire,
\IEEEmembership{Fellow, IEEE}%
\thanks{The authors are with the Communications and Information Theory Group, Technische Universit\"{a}t Berlin (\{m.barzegarkhalilsarai, saeid.haghighatshoar, caire\}@tu-berlin.de).}
}

\maketitle

\begin{abstract}
Massive Multiple-Input Multiple-Output (massive MIMO) is a variant of multi-user MIMO in which the number of antennas at each Base Station (BS) is very large and typically much larger than the number of users simultaneously served. Massive MIMO can be implemented with Time Division Duplexing (TDD) or Frequency Division Duplexing (FDD) operation. FDD massive MIMO systems are particularly desirable due to their implementation in current wireless networks and their efficiency in situations with symmetric traffic and delay-sensitive applications. However, implementing FDD massive MIMO systems is known to be challenging since it imposes a large feedback overhead in the Uplink (UL) to obtain channel state information for the Downlink (DL). In recent years, a considerable amount of research is dedicated to developing methods to reduce the feedback overhead in such systems. These studies focus on exploiting underlying channel structure such as low-rankness or sparsity in time, frequency, and space domains. In this paper, we use the sparse spatial scattering properties of the environment to achieve this goal. The idea is to estimate the support of the continuous, frequency-invariant scattering function from UL channel observations and use this estimate to obtain the support of the DL channel vector via appropriate interpolation. We use the resulting support estimate to design an efficient DL probing and UL feedback scheme in which the feedback dimension scales proportionally with the sparsity order of DL channel vectors. Since the sparsity order is much less than the number of BS antennas in almost all practically relevant scenarios, our method incurs much less feedback overhead compared with the currently proposed methods in the literature, such as those based on compressed-sensing. We use numerical simulations to assess the performance of our probing-feedback algorithm and compare it with these methods. \\
\textit{\textbf{keywords}: FDD massive MIMO systems, feedback overhead, sparse scattering function, support estimation, channel probing.}  

\end{abstract}

\section{Introduction}
The idea of using large antenna arrays at the \textit{Base Station} (BS), also known as \textit{massive Multiple-Input Multiple-Output} (massive MIMO) systems is proven to be promising for achieving very high-data rate connectivity in the next generation of mobile networks \cite{Marzetta-TWC10}. These systems provide major improvements with respect to the current technology in several aspects including increased data rate, enhanced reliability, energy efficiency, interference reduction, etc. \cite{larsson2014massive}. Implementing massive MIMO is much easier in \textit{Time Division Duplexing} (TDD) due to the inherent \textit{Uplink}-\textit{Downlink} (UL-DL) channel reciprocity \cite{marzetta2006much}. In this scenario, the channel state is obtained from the UL pilots transmitted from the users along orthogonal dimensions and is used to transmit/receive data to/from the users in the DL/UL via coherent beamforming. Unfortunately, the UL-DL channel reciprocity does not hold for \textit{Frequency Division Duplexing} (FDD) massive MIMO systems since the  UL and DL transmissions occur over disjoint frequency sub-bands. This implies that although the orthogonal pilot transmission is still necessary to obtain the channel state of the users in the UL to coherently beamform and receive UL data, the estimated UL channel state information can not be used to coherently beamform and transmit data to the users in the DL. This makes FDD massive MIMO schemes hard to implement. Unlike TDD systems, in FDD systems the BS needs to probe the channel in the DL and requests the users to feedback their channel state. This feedback overhead turns out to be overwhelming especially in massive MIMO systems where the number of BS antennas is large. Despite this issue, the FDD massive MIMO systems are still highly favorable because the current wireless networks are mostly based on FDD and FDD systems are more effective in situations with symmetric traffic and delay-sensitive applications \cite{jiang2015achievable,chan2006evolution,rao2014distributed}. As a result, in recent years, a significant effort is devoted to reduce the feedback overhead in these systems to make them practically feasible. 

In the general scheme for \textit{Channel State Information at the Transmitter} (CSIT) acquisition in FDD systems, the BS \textit{probes} the channel in the DL and the users upon receiving the transmitted pilots estimate their channel via estimators such as the least squares (LS) or the minimum mean squared error (MMSE) estimator. After this, the users feed back the estimated CSI to be used at the BS \cite{yin2013coordinated}. Numerous codebook based feedback methods have been proposed that carry out the task of CSI quantization. In these works the BS probes the channel, the users estimate their channel, quantize it according to a quantization codebook and feedback the corresponding code index to the BS. The feedback overhead in conventional codebook based methods scales linearly with the number of BS antennas $M$, i.e., ${\cal O} (M)$ \cite{love2003grassmannian,jindal2006mimo}, thus when $M\gg 1$, these methods impose dramatic overhead. The more recent codebook design techniques tend to reduce the overhead and can be classified in two major categories \cite{jiang2015achievable}: the designs based on time correlation of the channel vectors \cite{choi2015trellis,heath2009progressive,huang2009limited,ding2007multiple} and the designs based on spatial correlation of the channel vectors \cite{jiang2015achievable,adhikary2013joint,adhikary2014joint,nam2014fundamental}. We refer the interested reader to \cite{love2008overview} for an extensive overview of codebook based methods. 


A different type of CSI feedback method exploits low-rank or sparse channel models to reduce the overhead \cite{kuo2012compressive,sim2016compressed,rao2014distributed}. It exploits the fact that the channel vectors are sparse due to the local scattering environment between users and the BS, i.e., the signal received from a generic user at the BS consists of a few multi-path components with a limited \textit{Angle of Arrival} (AoA) support\footnote{Throughout the paper the term ``support" indicates a set of intervals/indices over which a function/vector has non-zero value.}. Compressed sensing techniques \cite{donoho2006compressed,candes2008introduction} are used to recover the whole channel vector at the user side by a handful of linear measurements (sketches) transmitted from the BS during the DL channel probing. In \cite{bajwa2010compressed}, the authors propose a compressed channel sensing and estimation method to obtain the impulse response of a frequency-selective channel and provide reconstruction error bounds. This work focuses on the sparsity in the delay domain. It is shown that efficient reconstruction is possible with the number of channel probings of the order ${\cal O}(s\log M)$, where $s$ is the channel sparsity level. In \cite{kuo2012compressive}, compressed channel feedback methods for spatially correlated channels are proposed by introducing a sparsifying dictionary (by adopting KLT) for the channel vector. Also, the random vector quantization (RVQ) and the Linde, Buzo, and Gray (LBG) algorithms are used for compressed CSI quantization. A dictionary-learning based approach for sparse channel modeling is presented in \cite{ding2016dictionary}. Exploiting angular UL-DL reciprocity, this work also proposes a joint UL-DL sparsifying dictionary which allows for compressed channel estimation with much less measurements. Another closely related compressed-sensing based CSI feedback method is presented in \cite{rao2014distributed}. In this work each user receives compressive measurements of its channel vector via training pilot symbols, feeds back these measurements to the BS and the whole channel estimation is done at the BS side. Assuming that all users share a part of their channel support as common support, the BS estimates the DL channel vectors of all the users by running a joint recovery algorithm, coined as \textit{Joint Orthogonal Matching Pursuit} (J-OMP). Exploiting this assumption in recovery, this technique further reduces the feedback overhead. 
\subsection{Contribution}
In this paper we focus on the spatial correlation of the channel vectors at the BS. This correlation is fully characterized by a continuous, frequency-invariant scattering function, which models the density of the power received from the user in the AoA domain. We make the key observation that although the channel vectors change independently across UL and DL, thus, the channel reciprocity does not hold, we still have a type of reciprocity since the second order statistic of the channel embedded in the scattering function is the same for the UL and the DL. We refer to this feature as the \textit{reciprocity of the scattering function}.  We use this structure to reduce the feedback overhead. Our contributions can be summarized as follows:
\begin{itemize}
	\item \textit{UL support estimation}: We use consecutive UL channel vectors received via UL pilot transmission and the joint sparsity of the channel vectors in the angular domain to estimate the support of the continuous scattering function. This step incurs almost no overhead since the UL channel vectors are naturally available at the BS as a result of UL pilot transmission.
	\item \textit{Exploiting the UL-DL reciprocity and interpolating the support}: By assuming the reciprocity of the scattering function, we interpolate the DL support from the support estimated in the UL. This steps incurs a support expansion that reduces the sparsity. However, when the support has a block structure, i.e. the non-zero elements appear in clusters, support interpolation is highly efficient. We should emphasize that this method is completely different from and generally simpler than previous compressed sensing methods in which sparsity assists recovery with less measurements. Here, we estimate the DL channel support without DL training. 
	\item \textit{Designing efficient probing vectors}: We propose an efficient design of probing vectors based on the estimated DL channel support. In particular, we propose a probing scheme that exploits the common support among users to obtain a better estimate of the corresponding channel coefficients.
	\item \textit{Reducing feedback overhead}: The users send back the channel measurements to the BS as in \cite{rao2014distributed}. Using the estimated DL channel support, the BS estimates the channel vector with much fewer measurements compared to the conventional compressed sensing methods and channel estimation reduces to a simple and fast Least Squares (LS) estimation. 
\end{itemize}

Our simulation results show that the proposed method outperforms the compressed-sensing based approach in terms of system spectral efficiency and reduced feedback overhead.

\subsection{Notations}
We denote vectors by boldface small letters (e.g. $\bfx$), matrices by boldface capital letters (e.g. $\bfX$), scalars by non-boldface letters (e.g. $x$ or $X$), and sets by calligraphic letters (e.g. $\mathcal{X}$). The $i$\textsuperscript{th} element of a vector $\bfx$ and the $(i,j)$\textsuperscript{th} element of a matrix $\bfX$ will be denoted by $[{\bfx}]_i$ and $[{\bfX}]_{i,j}$. For a matrix $\bfX$, we denote its $i$\textsuperscript{th} row and $j$\textsuperscript{th} column with the row vector $\bfX_{i,.}$ and the column vector $\bfX_{.,j}$, respectively. For convenience, the index of the entries of a vector starts from $0$, i.e., $[\bfx]_0$ denotes the first entry of vector $\bfx$. We denote the Hermitian and the transpose of a matrix $\bfX$ by $\bfX^\herm$ and $\bfX^\transp$, respectively, with the same notation being used for vectors. We use $\Vert \bfx\Vert$ for the $\ell_2$-norm of a vector $\bfx$, and $\Vert \bfX\Vert= \langle \bfX, \bfX \rangle $ for the Frobenius norm of a matrix $\bfX$. We always denote the identity matrix of order $p$ with $\bfI_p$. For arguments that are intervals over the real line, $|\cdot|$ returns the length of the interval and for arguments that are discrete sets, it returns the cardinality of the set. For an integer $k$, we use the shorthand notation $[k]$ to denote the set of integers $\{0,1, . . . , k-1\}$.
\section{System Setup}\label{sec:sys_setup}
\subsection{Channel Model}
\begin{figure}[t]
	\centering
	\includegraphics{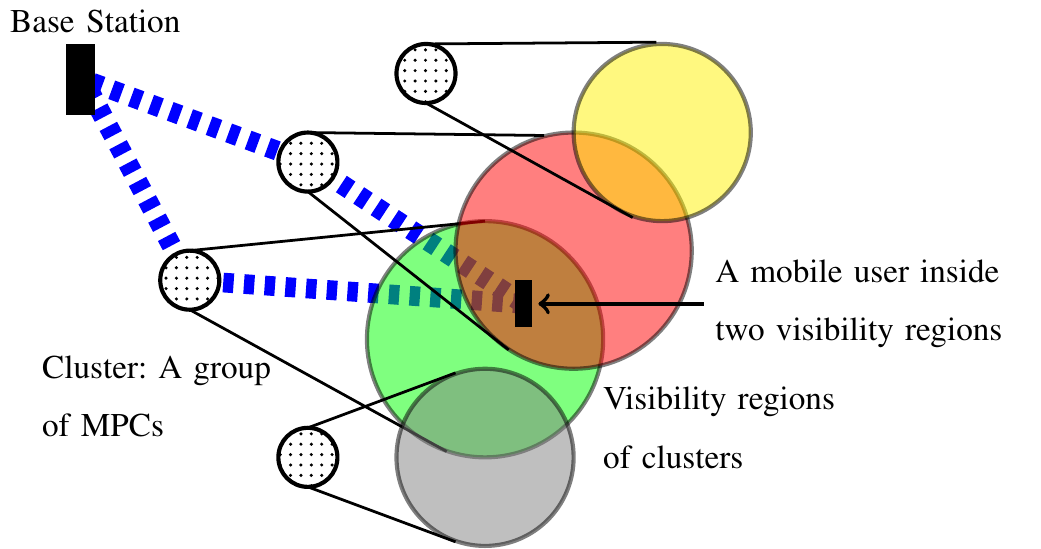}
	\caption{A sketch of the cluster and visibility regions of the COST2100 model.}
	\label{COST2100}
\end{figure}

We consider the COST 2100 channel model as the basic setup for modeling a propagation environment \cite{liu2012cost}. This model is a geometry-based stochastic channel model (GSCM) that describes the properties of the channel in time, frequency and space. The propagation model consists of clusters of \textit{Multipath Components} (MPCs) and visibility regions as its building blocks. A cluster is a group of MPCs, generated by the reflection of the signal from the objects in the environment. A visibility region is a region corresponding to only one cluster and determines the area over which the \textit{user equipment} (UE) can connect to the BS through that particular cluster. In practice, the UE might move inside an intersection of visibility regions, hence connecting to the BS through several clusters. Fig. \ref{COST2100} shows an scheme of the COST 2100 propagation model.

  This model implies that the scattering geometry of the channel between the BS antenna array and the UE antenna array 
 remains constant over time intervals corresponding to the UE remaining in the same intersection of visibility regions. 
 In contrast, in correspondence to the motion of the UE such that such intersection of visibility regions changes, we have \textit{sharp} 
 (generally discontinuous) transitions of the scattering geometry. 
 Since moving across the regions occurs at a time scale much larger than moving across one wavelength (specially at mm-Wave lengths), 
 it is safe to assume that the channel scattering geometry is piecewise time-invariant. Here we focus on this piece-wise stationary situation and 
 consider the channel model for a given {\em fixed scattering geometry}. We also incorporate the well-known \textit{Wide Sense Stationary Uncorrelated Scattering} (WSSUS) assumption which states that at any time instant the channel gains of different signals paths are uncorrelated.

\subsection{Array and Signaling Model}
\begin{figure}[t]
	\centering
	\includegraphics{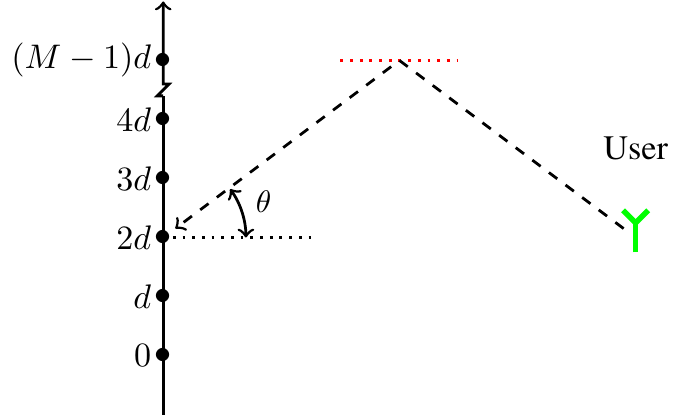}
	\caption{{\small 
			Array configuration in a multi-antenna receiver in the presence of a single scatterer with an angle of arrival $\theta$.}}
	\label{fig:sc_channel}
	\end{figure}

Consider a BS with a uniform linear array (ULA) with $M \gg 1$ antennas and a UE with a single antenna. 
The geometry of the array is shown in Fig. \ref{fig:sc_channel}. This figure illustrates the coordinates of the antenna elements of the BS array, and how the AoA $\theta$ is measured. With such arrangement, the coordinates of the $i$\textsuperscript{th} BS antenna are denoted as $(0,id)$, for $i\in [M]$, where $d$ denotes the spacing between two consecutive antennas.

We also assume that UEs and the BS use OFDM signaling, both in UL and DL, although in different frequency bands. Since stationarity holds along subcarriers, in the following we focus on the communication over a single subcarrier to simplify notation and address the subtleties caused by considering the OFDM signaling where necessary. In the next section, we develop a complex-valued channel model for the physical channel in both UL and DL. The goal is to show the connections between the UL and DL channels, to be later exploited by our DL channel sensing and estimation method.  

\subsection{Uplink Channel}
\begin{figure}[t]
		\centering
	\includegraphics{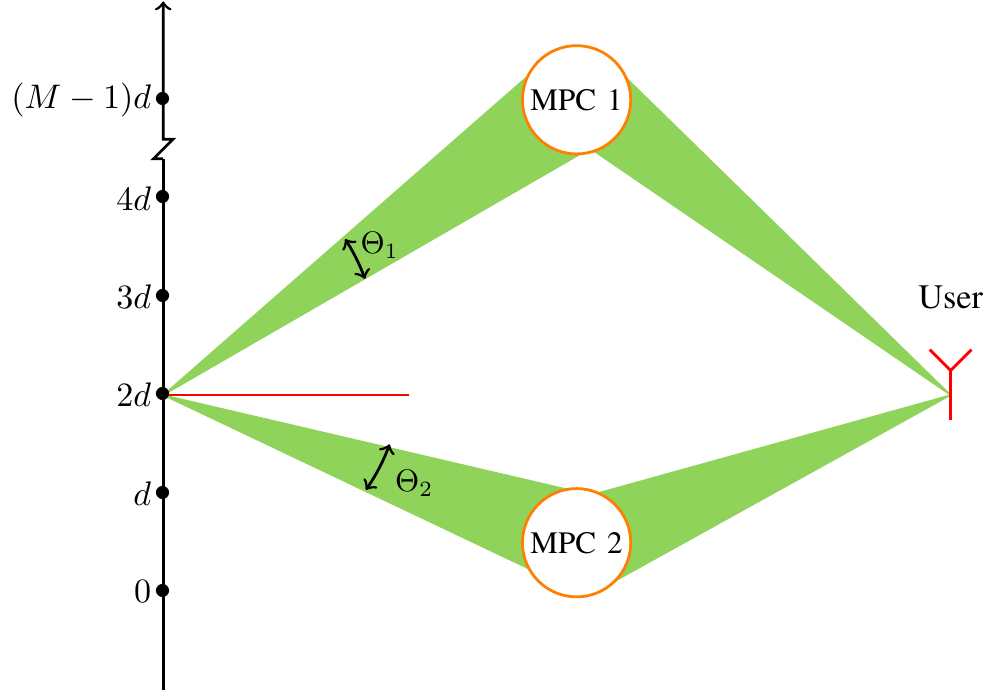}
		\caption{An example of the angular support, generated by two MPCs. Here the support of the scattering is confined to the intervals $\Theta_1$ and $\Theta_2$.}
		\label{ang_intervals}
	\end{figure}

\begin{figure}[t]
	\centering
	\includegraphics{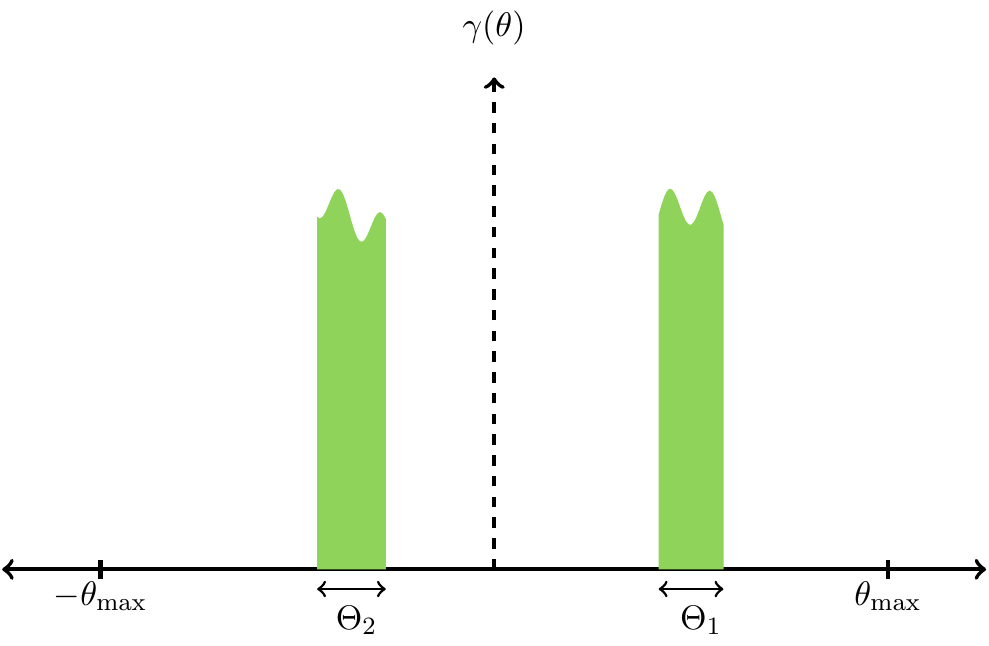}
	\caption{The scattering function corresponding to Fig. \ref{ang_intervals}}
	\label{example_scat_func}
\end{figure}

 We consider a general scattering model in which the received signal of the user comes from a continuum of AoAs, similar to the one proposed in \cite{haghighatshoar2015channel}. One snapshot of the received signal during pilot transmission is given by
\begin{equation}\label{rec_sig}
\bfr(t) = \bfh_{\rm ul}(t) x(t) + \bfn(t) := \int_{\Theta} \rho_{\rm ul}(\theta;t) \aulvec  {\rm d}\theta ~x(t)~  + \bfn(t),
\end{equation}
where $\bfh_{\rm ul}(t):=\int_{\Theta} \rho_{\rm ul}(\theta;t) \aulvec {\rm d}\theta \in \bC^M$ denotes the channel vector of the user, where $\Theta:=[-\theta_{\max}, \theta_{\max})$ is the angular range scanned by the BS array, where $x(t)\in \bC$ is the transmitted UL pilot symbol of the user along the channel vector $\bfh_{\rm ul}(t)$, which typically belongs to a signal constellation such as QAM, where $\bfn(t) \sim \cg({\bf 0}, \sigma^2\Id_M)$ is the Additive White Gaussian Noise (AWGN) of the  antenna elements, and where $\aulvec \in \bC^M$ is the UL array response at AoA $\theta$, whose $i$\textsuperscript{th} component is given by 
\begin{equation}\label{eq:arr_resp_ul}
[\bfa_{\rm ul}(\theta) ]_i= e^{j \frac{2\pi}{\lambdaul} i d \sin\theta}. \\
\end{equation}
In \eqref{eq:arr_resp_ul}, $\lambdaul=\frac{c}{f_{{\rm ul}}}$ is the carrier wave length over UL frequency band, where $c$ is the speed of light and $f_{{\rm ul}}$ is the carrier frequency in the UL frequency band. In \eqref{rec_sig}, $\rho_{\rm ul}(\theta;t) $ denotes a complex circularly symmetric Gaussian random process representing the random gain of the scatterers at different AoAs. This random process is completely characterized by its second order statistics. Assuming
mean zero, i.e., $\EE[\rho_{\rm ul}(\theta;t)]=0$, and the WSSUS model we have
\begin{equation} 
\EE[\rho_{\rm ul}(\theta;t) \rho_{\rm ul}^*(\theta';t)]  = \gamma(\theta) \delta(\theta - \theta'),
\end{equation}
where $\gamma(\theta)$ is the AoA {\em scattering function}, which represents the received signal energy density as a function of the AoA. As described before, propagation takes place through clusters of MPCs so that the energy density is concentrated on a very limited AoA support and the scattering function $\gamma (\theta)$ is sparse. When $M \gg 1$, this translates into channel vectors that are sparse in the angular domain. Fig. \ref{ang_intervals} illustrates such a sparse propagation model with two MPCs with a limited angular support. The corresponding scattering function is depicted in Fig. \ref{example_scat_func}. We further elaborate on the sparse channel model in the following. 	
\subsection{Description of Uplink Channel in Fourier Basis}\label{sec:fourier_description}
In this section, for convenience, we first introduce a finite-dimensional representation of the array response by quantizing the range of AoAs. To this purpose, consider the $M \times M$ unitary Discrete Fourier Transform (DFT) matrix 
	\begin{equation} \label{IDFT_mat}
	\left[\Fm\right]_{k,\ell} := 	\tfrac{1}{\sqrt{M}} {\rm e}^{{\rm j} \frac{2\pi}{M}k (\ell -\frac{M}{2})},
	\end{equation}
	with $k=0,\ldots,M-1 $ and $ l=0,\ldots,M-1$. The matrix $\Fm$ forms a unitary basis for $\CC^M$; Therefore, any vector $\av$ can be represented as a linear combination of the columns of $\Fm$ as
	\[ \av = \Fm \check{\av} = \sum_{i=0}^{M-1}[\check{\av}]_i \Fm_{.,i} ~. \]
	The vector $\check{\av}$ denotes the representation of $\av$ in DFT basis and its $i$\textsuperscript{th} component is given by $[\check{\av}]_i = \Fm_{.,i}^\herm \av$. In particular, for the UL array response we define
	\begin{eqnarray}
	[\aulveccheck]_i  & := & \Fm_{.,i}^\herm \aulvec \\
	& = & \frac{1}{\sqrt{M}} \sum_{\ell = 0}^{M-1} e^{-j\frac{2\pi}{M} \ell (i-\frac{M}{2})} e^{j2\pi \frac{d}{\lambdaul} \ell \sin \theta} \\
	& = & \frac{1}{\sqrt{M}} \frac{1 - e^{j2\pi (\frac{d}{\lambdaul} \sin \theta - \frac{i}{M}+\frac{1}{2} ) M}}{1 - e^{j2\pi (\frac{d}{\lambdaul} \sin \theta - \frac{i}{M} +\frac{1}{2} )}} \\
	& = & \frac{1}{\sqrt{M}} \frac{\sin \left ( \pi ( \frac{d}{\lambdaul} \sin \theta - \frac{i}{M}+\frac{1}{2} ) M \right )}{\sin \left ( \pi (\frac{d}{\lambdaul} \sin \theta - \frac{i}{M}+\frac{1}{2} ) \right )}  
	e^{j\pi ( \frac{d}{\lambdaul} \sin \theta - \frac{i}{M}+\frac{1}{2}) (M-1)}.
	\end{eqnarray}
	The function $D_M(\psi)=\frac{\sin(\pi \psi M)}{\sin(\pi \psi)}$ is the Dirichlet kernel with parameter $M$. The Dirichlet kernel has a peak at $\psi=0$ and has non-negligible 
	magnitude only for $|\psi| \leq 1/M$, where $1/M$ reflects the spatial resolvability of the ULA. It follows that the coefficients $[\aulveccheck]_i$ are 
	significant only when $\left| \psi \right| = \left|\frac{d}{\lambdaul} \sin \theta - \frac{i}{M}+\frac{1}{2}\right| \le 1/M$. Let $\bfh_{\rm ul}(t)$ be the channel vector of the user at time slot $t$ as before. We denote the representation of this channel vector in the DFT basis by $\bfh_{\rm ul}(t) = \bfF \check{\bfh}_{\rm ul}(t)$, where
	\begin{equation}
	\check{\bfh}_{\rm ul}(t) := \int_{\Theta} \rho_{\rm ul}(\theta;t) \aulveccheck  {\rm d}\theta,
	\end{equation}
	Now, consider the $i$\textsuperscript{th} element of the vector $\check{\bfh}_{\rm ul}(t)$, which is a random variable with mean $\bE\{[\check{\bfh}_{\rm ul}(t)]_i\} = \int_{\Theta} \bE\left\{\rho_{\rm ul}(\theta;t)\right\} \left[\aulveccheck\right]_i  {\rm d}\theta = 0$ and variance
	\begin{eqnarray}
	\bE\left\{\left| [\check{\bfh}_{\rm ul}(t)]_i\right|^2\right\} & = & 	\bE\left\{[\check{\bfh}_{\rm ul}(t)]_i [\check{\bfh}_{\rm ul}(t)]_i^\ast\right\} \\
	& = & \int_{\Theta'} \int_{\Theta} \bE\{\rho_{\rm ul}(\theta;t) \rho_{\rm ul}(\theta';t)^\ast \} \left[\aulveccheck\right]_i  \left[\aulveccheck\right]_i ^\ast   {\rm d}\theta {\rm d}\theta' \\
	& = & \int_{\Theta'} \int_{\Theta} \gamma(\theta) \delta (\theta - \theta')   \left[\aulveccheck\right]_i  \left[\aulveccheck\right]_i ^\ast  {\rm d}\theta {\rm d}\theta' \\
		& = &\int_{\Theta} \gamma(\theta)  \left| \left[\aulveccheck\right]_i  \right|^2 {\rm d}\theta\\
	& = & \frac{1}{M}  \int_{\Theta}   \gamma(\theta)    \left|       
	\frac{\sin \left ( \pi ( \frac{d}{\lambdaul} \sin \theta - \frac{i}{M}+\frac{1}{2} ) M \right )}{\sin \left ( \pi (\frac{d}{\lambdaul} \sin \theta - \frac{i}{M}+\frac{1}{2} ) \right )} \right|^2  {\rm d}\theta \\
		& = & \frac{1}{M}  \int_{\Theta}   \gamma(\theta)    \left|       
		\frac{\sin \left ( \pi \psi_{{\rm ul},i}(\theta) M \right )}{\sin \left ( \pi \psi_{{\rm ul},i} (\theta)\right )} \right|^2  {\rm d}\theta \label{eq_last},
	\end{eqnarray}
	where we have defined $\psi_{{\rm ul},i}(\theta) := \frac{d}{\lambdaul} \sin \theta - \frac{i}{M}+\frac{1}{2}$. As described before, $ \left| \left[\aulveccheck\right]_i  \right|^2 = \left|       
	\frac{\sin \left ( \pi \psi_{{\rm ul},i}(\theta) M \right )}{\sin \left ( \pi \psi_{{\rm ul},i}(\theta) \right )} \right|^2$ has a significant magnitude only for $\theta \in \xiul$, where
	\begin{equation}
	\xiul	=\{\theta ~|~ |\psi_{{\rm ul},i}(\theta)| \leq \frac{1}{M} \}.
	\end{equation}
 Let $\Xgam$ denote the support of the scattering function $\gamma(\theta)$, i.e.,
	\begin{equation}
	\Xgam:= \{\theta ~|~ \gamma(\theta) \neq 0\}.
	\end{equation}
	From the discussion above and \eqref{eq_last} it is obvious that $[\check{\bfh}_{\rm ul}(t)]_i$ has significant power only if the interval $\xiul$ has a non-empty intersection with $\Xgam$, i.e., $\xiul \cap \Xgam \neq \emptyset$. In particular, we have the following observation.
	
\begin{obs_env}\label{env:obs_1}
     \textit{The length of the interval $\xiul$ is given by
     \begin{equation}
     |\xiul| = \left| \sin^{-1}\left( \frac{\lambdaul}{d} \left(\frac{i+1}{M} -\frac{1}{2}\right) \right) - \sin^{-1}\left( \frac{\lambdaul}{d} \left(\frac{i-1}{M} -\frac{1}{2}\right) \right) \right|.
     \end{equation}
     This shows that for a reasonable choice of the antenna spacing $d$, we have that $|\xiul| \approx \Oc (1/M)$. Therefore, $[\check{\bfh}_{\rm ul}(t)]_i$ has significant variance only for a small set of indices $i$. We denote this set by $\suppul$ and define it as
          \begin{equation}
          \suppul = \{ i\in [M] ~|~ \xiul \cap \Xgam \neq \emptyset \}.
          \end{equation} 
     	     It is obvious that $|\suppul | \ll M$, implying that $\check{\bfh}_{\rm ul}(t)$ is a sparse vector. Furthermore, with $M\gg 1$ for each interval of the support $\mathcal{X}_{\gamma}$ corresponding to an MPC, there exists a block of non-zero channel coefficients.   }
\end{obs_env}

Observation \ref{env:obs_1} draws a connection between the sparse scattering function $\gamma(\theta)$ and the block-sparse channel coefficients vector $\check{\bfh}_{\rm ul}(t)$. The block-sparse structure helps us both in estimating the support from sub-sampled sketches of the channel vector and obtaining an accurate estimation of the DL channel support. In the next section we describe DL transmission, with special focus on how the UL and DL support sets relate to each other. 

\subsection{Downlink Channel}
Similar to the UL channel, the DL channel can be formulated as
\begin{equation}\label{dl_channel}
\bfh_{\rm dl}(t) = \int_{\Theta} \rho_{\rm dl}(\theta ; t) \adlvec {\rm d}\theta  \in \bC^M,
\end{equation}
where $\adlvec \in \bC^M$ is the DL array response at AoA $\theta$, whose $i$\textsuperscript{th} component is given by 
\begin{equation}
[\adlvec]_i= e^{j \frac{2\pi}{\lambdadl} i d \sin\theta}, \\
\end{equation}
where $\lambdadl=\frac{c}{f_{{\rm dl}}}$ and $f_{{\rm dl}}$ are the carrier wavelength and the carrier frequency in the DL frequency band, respectively. In \eqref{dl_channel}, $\rho_{\rm dl}(\theta;t) $ denotes a complex circularly symmetric Gaussian random process as before with zero mean and the same correlation function
\begin{equation}
\EE[\rho_{\rm dl}(\theta;t) \rho_{\rm dl}^*(\theta';t)]  = \gamma(\theta) \delta(\theta - \theta').
\end{equation}
The processes $\rho_{\rm ul}(\theta;t) $ and $\rho_{\rm dl}(\theta;t) $ are essentially independent since the multipath gains are not correlated beyond the coherence frequency. However, they share almost the same power spread profile, denoted by the scattering function $\gamma (\theta)$, since it only depends on geometrical properties of the propagation environment and not on the carrier frequency. As a result, although channel reciprocity does not hold here, we can leverage the reciprocity of the scattering function to draw a connection between channel support over UL and DL. We do this by representing the DL channel in Fourier basis, similar to its UL counterpart as described in section \ref{sec:fourier_description}.
	\subsection{The Connection between Downlink and Uplink Support Sets}\label{sec:supp_connection}
Consider the DFT matrix defined in \eqref{IDFT_mat}. Following the analysis of the previous section, we have
\begin{eqnarray}
	\bfh_{\rm dl}(t)= \bfF \check{\bfh}_{\rm dl}(t), 
\end{eqnarray}
where $\check{\bfh}_{\rm dl}(t) $ represents the vector of DL channel coefficients in DFT basis. The $i$\textsuperscript{th} element of $\check{\bfh}_{\rm dl}(t) $ is a random variable with a zero mean and a variance given by
	\begin{eqnarray}
	\bE\{| [\check{\bfh}_{\rm dl}(t)]_i|^2\} & = & 	\bE\{[\check{\bfh}_{\rm dl}(t)]_i [\check{\bfh}_{\rm dl}(t)]_i^\ast (t)\} \\
			& = & \frac{1}{M}  \int_{\Theta}   \gamma(\theta)    \left|       
			\frac{\sin \left ( \pi \psi_{{\rm dl},i}(\theta) M \right )}{\sin \left ( \pi \psi_{{\rm dl},i} (\theta)\right )} \right|^2  {\rm d}\theta,
	\end{eqnarray}
	where $\psi_{{\rm dl},i}(\theta) := \frac{d}{\lambdadl} \sin \theta - \frac{i}{M}+\frac{1}{2}$. Similar to the case of UL transmission, we define $\xidl$ to be an interval on the real line over which  $\left| \left[\adlveccheck\right]_i  \right|^2 $ has significant magnitude. This interval is defined as $ \xidl =\{\theta ~|~ |\psi_{{\rm dl},i}(\theta)| \leq \frac{1}{M} \}$. As before, the element $[\check{\bfh}_{\rm dl}(t)]_i$ has significant variance only if the interval $\xidl$ has a non-empty intersection with $\Xgam$, i.e., $\xidl\cap \Xgam \neq \emptyset$. Let $\suppdl$ be the set of indices over which $[\check{\bfh}_{\rm dl}(t)]_i$ has significant variance. This set is defined by
	\begin{equation}\label{dl_supp_1}
	\suppdl := \{ i\in [M] ~|~ \xidl\cap \Xgam \neq \emptyset \}.
	\end{equation} 
	Now, suppose that we have access to the UL support set 	$\suppul$ and we want to check whether one can specify the DL support set $	\suppdl$ given this information or not. To do so, note that the UL support set determines the set of intervals $\xiul$ that have a non-empty intersection with $\Xgam$. In addition, it is easy to show that the set $\{\xiul\}_{i=0}^{M-1}$ covers the angular domain $\Theta$. Hence, we conclude that
	\begin{equation}
	\Xgam \subseteq \Xgamhat:= \underset{i\in \suppul}{\Bcup} \xiul.
	\end{equation}
 In other words, the original angular support of the scattering function is covered by the intervals $\xiul$. As the number of BS antennas $M$ increases (as is the case in the massive MIMO scenario), the number of such intervals increases while the length of each interval decreases. This means that with increasing $M$ the angular domain and in particular $\Xgam$ will be finer covered by the set of intervals $\xiul$. Therefore, by knowing $\suppul$ one can assure a tight localization of the support of the scattering function, i.e., $\Xgam$. 
\begin{figure}[t]
	\centering
\includegraphics{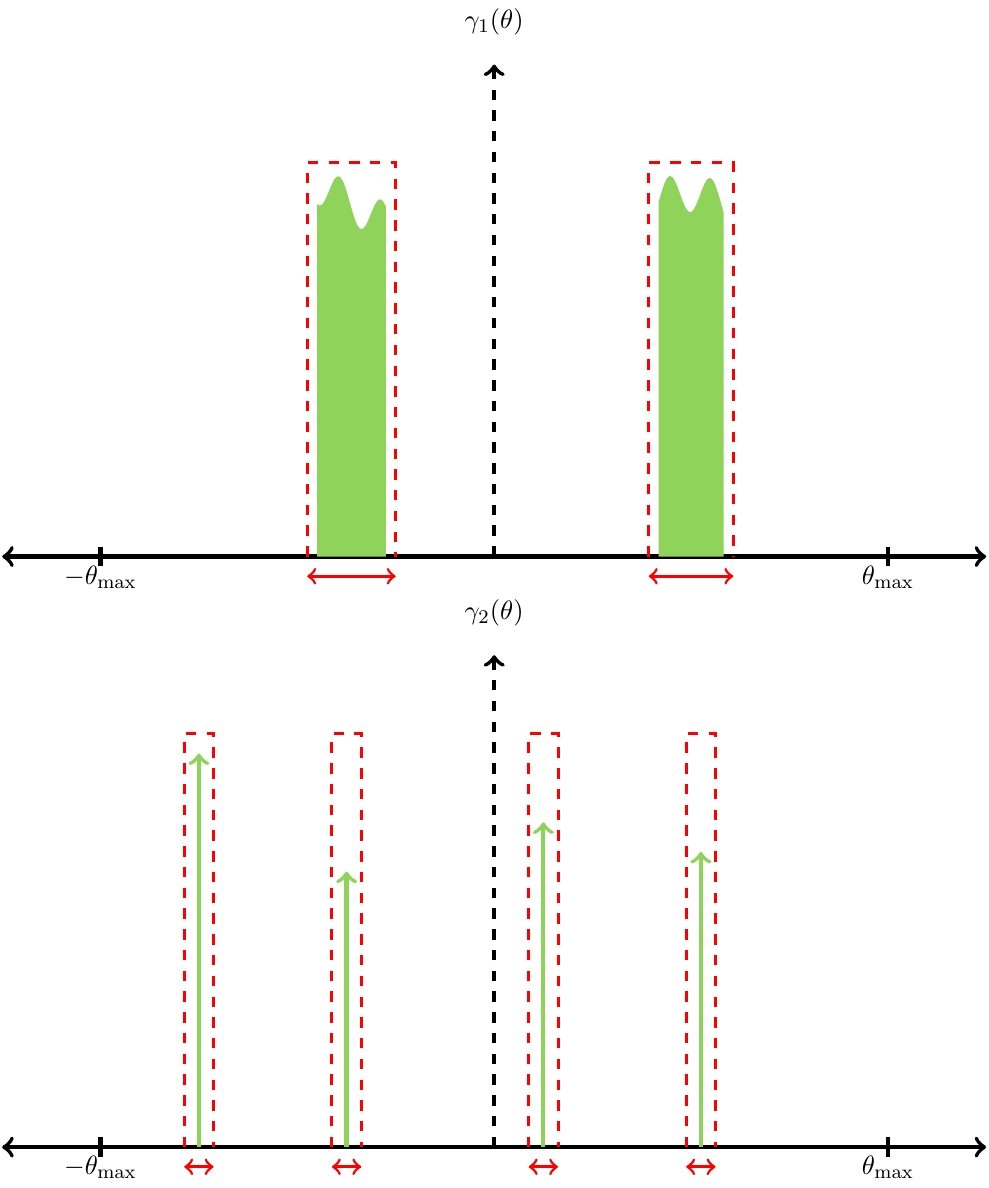}
	\caption{Comparison of support estimation for two different scattering functions $\gamma_1 (\theta)$ and $\gamma_2(\theta)$. The dashed red lines specify the estimated support for each of these functions. The first scattering function has a block-sparse structure and the second one has sparse but not block-sparse structure with specular components.}
	\label{compare_block_structure}
\end{figure}
	We can estimate the DL support set $\suppdl$ as follows
	\begin{equation}
	\suppdlhat = \{ i\in [M] ~|~ \xidl\cap \Xgamhat \neq \emptyset \},
	\end{equation}
which means that we obtain an estimate of the DL support by looking at the estimated support of the continuous scattering function $\gamma (\theta)$. Note that the block-sparsity of the channel vectors is particularly helpful here, because it results in a limited over-estimation of the support compared with the case where a block-sparse structure does not exist. This fact is illustrated in Fig. \ref{compare_block_structure}. In this figure we have schematically compared support estimation for two different scattering functions, where one of them has a block-sparse structure and the other one has a sparse but not block-sparse structure. When the scattering function is block-sparse (the above plot in Fig. \ref{compare_block_structure}), support over-estimation is hardly an issue. However, when the scattering function is not block-sparse (the below plot in Fig. \ref{compare_block_structure}), over-estimation of the support is substantial. Therefore, when a block structure is present, as is the case in many Massive MIMO channel models \cite{poutanen2010significance}, the support estimation method developed in this section has a good performance. 

	\section{Downlink support estimation from uplink observations}\label{sec:dl_supp_from_ul_supp}

A main ingredient of our method is estimating the support of the scattering function in the UL. In this section, we propose an algorithm that estimates the UL channel and thereby its angular support. We consider a general case where the number of RF chains $m$ at the BS is much less than the number of antennas $M$, meaning that instead of the UL channel vector, only an $m$-dimensional sketch of it is available at the BS during UL pilot transmission phase. This constraint is sometimes considered in massive MIMO literature due to the practical limitation on the number of RF chains or limited analog to digital conversion rate \cite{haghighatshoar2016low}. The low-dimensional sketch of the channel vector is obtained by an $m\times M$ projection matrix. Since we are estimating the signal subspace (support of the scattering function), we can use several UL channel vector realizations, which can be obtained across several subcarriers (via the pilots transmitted from the users) due to the stationarity in frequency or even across several OFDM symbols due to the stationarity in time.

		\subsection{Uplink Support Estimation}\label{sec:up_supp_es}

	Let $\bfB \in \bC^{m\times M}$ be the projection matrix introduced above. Here we consider a particularly simple \textit{antenna selection} scheme in which $\bfB$ is a binary 0-1 selection matrix with a single $1$, randomly located in each row and the locations are distinct across all rows. Using this projection matrix is equivalent to sampling a subset of size $m\ll M$ of antenna elements, while receiving the signal in UL. From a compressed sensing point of view, this is a good choice for a projection matrix because the channel vectors are sparse in the DFT basis and antenna sampling is incoherent with the DFT basis \cite{foucart2013mathematical}. 

	Let assume that during UL transmission each user sends $L$ symbols $\{x_i(t)\}_{i=1}^L$ through $L$ frequency subcarriers. Without loss of generality we can assume $x_i(t)=1$ for all $i$. We denote by $\Kc$ the set of selected antenna indices. The low-dimensional noisy projection of the received signal at the BS in subcarrier $i$ can be written as
	\begin{equation}
	\yv_i(t) = \Bm \,  \hv_{{\rm ul},i}(t) + \nv_i(t),
	\end{equation}
	where $\hv_{{\rm ul},i}(t)$ is the channel vector corresponding to the $i$\textsuperscript{th} subcarrier and $\nv_i(t)\sim \cg({\bf 0}, \sigma^2\Id_m)$ is the AWGN. Let $\tilde{\Fm} \in \bC^{M\times qM}$ denote the overcomplete DFT dictionary with oversampling factor $q$, defined as
    \begin{equation}
    \left[\tilde{\Fm}\right]_{k,\ell} := 	\tfrac{1}{\sqrt{M}} {\rm e}^{{\rm j} \frac{2\pi}{qM}k (\ell -\frac{qM}{2})},
     \end{equation}	
	with $k =0, \ldots, M-1 $ and $\ell = 0, \ldots, qM-1$. We use the overcomplete DFT matrix as the sparsifying dictionary for the purpose of support estimation. This gives us more freedom in estimating the support of the scattering function, because now consecutive entries of the sparse coefficients vector overlap and their centers are $1/qM$ apart, while this spacing is equal to $1/M$ when we use a DFT matrix\footnote{We use the overcomplete DFT only for the purpose of UL support estimation.}. Now, the vectors $\left\{ \hv_{{\rm ul},i}(t) \right\}_{i=1}^L$ have a common support over the dictionary $\tilde{\Fm} $, since we assume that the frequency variation over the UL band is negligible. The problem then is to estimate this common support from a set of noisy incomplete observations. Define $\Ym = \left[\yv_1(t), \ldots, 	\yv_L(t)\right]$ and $\Nm=\left[\nv_1(t),\ldots, \nv_L(t)\right]$. The support estimation problem amounts to finding the set of indices corresponding to the non-zero rows of the solution matrix $\hat{\Xm}\in \bC^{qM\times L}$ in a \text{Multiple Measurement Vectors} (MMV) problem. This problem can be formulated as follows,
	\begin{equation}\label{l2_1_first_formula}
	\begin{aligned}
	\hat{\Xm}\, = \,	& \underset{\Xm \in \bC^{qM\times L}}{\arg \min}&& \Vert \Xm\Vert_{2,1}, \\
		& \text{subject to} && \Vert \Ym - \Gm \Xm\Vert \le \sqrt{mL} ~\sigma, \\
	\end{aligned}
	\end{equation}
	where $\Gm := \Bm \tilde{\Fm}\in \bC^{m\times qM}$ and the $\ell_{2,1}$-norm is defined by $\Vert \Xm \Vert_{2,1}  = \sum_{i=0}^{L-1} \Vert \Xm_{i,.}\Vert_2$. To solve this problem we make use of the low-complexity algorithm proposed in \cite{haghighatshoar2017massive} for subspace estimation. Once \eqref{l2_1_first_formula} is solved, we obtain the support by calculating the $\ell_2$-norm of each row of matrix $\Xm$. If the $\ell_2$-norm of a particular row is greater than a certain threshold, then it is labeled as \textit{active} and otherwise it is labeled as \textit{inactive}. In other words
	\begin{equation}
	\hat{\Sc}_{\rm ul} := \{i\in [M] ~|~ \Vert \hat{\Xm}_{i,.}\Vert_2 \ge \epsilon \}
	\end{equation}
	 denotes the estimated UL channel support for one user. The threshold $\epsilon$ can be set according to the available information about the sparsity level of the channel coefficients, which can be acquired over time. In addition, our algorithm shows a fast convergence with oversampling factors $q=2$ or $q=3$.
	 
 	\subsection{Downlink Support Estimation}
\begin{figure}[t]
	\centering
	\includegraphics{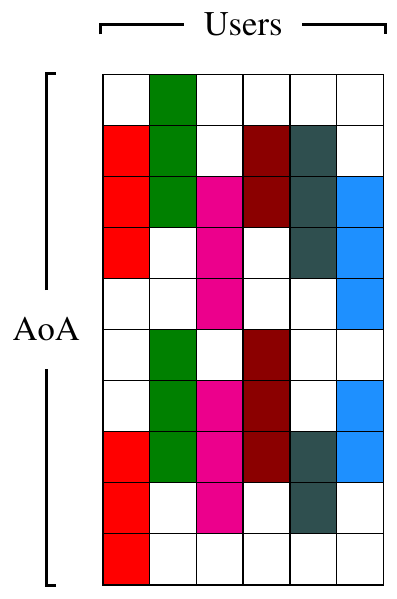}
	\caption{Schematic of DL support profile for different users, available at the BS before DL probing. The colored grid points represent support elements.}
	\label{supp_sets}
\end{figure}

 	The DL support can be estimated from the estimated UL support using the ideas developed in section \ref{sec:fourier_description}. To this purpose, we first estimate the angular support of the scattering function as
 	\begin{equation}\label{eq:Gamma_est_1}
 	\Xgamhat = \underset{j\in \suppulhat}{\Bcup} \Ic_{{\rm ul}, j} \, .
 	\end{equation}
This gives a fine approximation of the support of the scattering function, particularly when it has a block-sparse structure and $M\gg 1$ , as described in section \ref{sec:supp_connection}. Now, similar to \eqref{dl_supp_1}, the DL support is estimated by
	\begin{equation}\label{eq:dl_supp_2}
	\suppdlhat = \{ i\in [M] ~|~ \xidl\cap \Xgamhat \neq \emptyset \}.
	\end{equation} 
	This means that we determine those indices whose corresponding intervals intersect with the estimated support of the continuous scattering function. Fig. \ref{supp_sets}  illustrates the estimated DL support profile for different users. The BS leverages this information in order to effectively probe the DL channel and to reduce feedback overhead. 
	\section{Downlink Channel Probing and Estimation}\label{sec:sensing_and_estimation}

In this section, by using the estimated DL channel support, we propose a scheme that substantially reduces the feedback overhead. The proposed feedback scheme promises an overhead that grows linearly only with respect to the number of resolvable angular paths or equivalently the support size of the DL channel. To estimate the DL channel, the BS broadcasts $T$ probing vectors in $T$ consecutive time slots. We denote the transmitted probing vector in the $j$\textsuperscript{th} time slot by ${\boldsymbol \phi}_j \in \bC^{M\times 1}$, and denote the set of BS probing vectors by a matrix ${\bf\Phi}  \in \bC^{T\times M}$ where ${\bf\Phi}_{j,.}={\boldsymbol \phi}_j^\transp$. The received signal at user $i$ after $T$ time slots can be written as 
	 \begin{equation}\label{eq:cs_eq_1}
	 {\bfy}^{(i)}= {\bf\Phi} {\bfh}_{{\rm dl}}^{(i)}(t) + {\bfn}^{(i)},
	 \end{equation}
	 where $\bfh_{{\rm dl}}^{(i)}(t)$ is the DL channel vector for user $i$ and $\bfn^{(i)} \sim \cg({\bf 0},{\bf I}_T)$ is the AWGN vector at the user side, with i.i.d unit-variance entries. We assume that $\Vert {\boldsymbol \phi}_j\Vert^2=\Pbf$ for all $j$, where $\Pbf$ is the power spent on a single probing vector by the BS. Also notice that DL beam transmission is similarly carried out across multiple DL subcarriers, but we focus on a single subcarrier and drop subcarrier index to simplify the notation. 
	 
	 Equation \eqref{eq:cs_eq_1} resembles a compressed sensing problem in which $\bfy^{(i)}$ is the measurements vector, $\bfB$ is the sensing matrix and $\bfh_{{\rm dl}}^{(i)}(t)$ is a vector that is sparse in the DFT domain. As in a compressed sensing problem, our goal is to reduce the number of measurements, since it will result in reduced feedback overhead.  This reduction is particularly interesting in our setting, since an estimation of the DL support is available at the BS and one can estimate the channel with much fewer measurements.
	 
	 \subsection{Channel Probing Schemes}\label{subsec:sensing_schemes}

	 First, we set the number of probing vectors to be equal to the maximum estimated DL support size among all users, i.e., $T=\underset{i\in [K]}{\max} ~|\suppdlhat^{(i)}|$, where $\suppdlhat^{(i)}$ is the estimated DL support for user $i$. We introduce three probing schemes, that the BS can use to probe the DL channel.
	 \begin{itemize}
	 	\item Gaussian Probing\\
	 	In this scheme the probing vectors are generated by a complex Gaussian distribution, i.e., ${\bphi}_j \sim \cg({\bf0},{\bfI}_M)$ for $j=1,\ldots,T$. Then we normalize these vectors so that $\Vert {\bphi}_i\Vert^2 = \Pbf$ for all $i$. Now, transmitting such probing vectors is similar to taking Gaussian measurements from a sparse vector in the literature of compressed sensing.  In \cite{rao2014distributed}, the Gaussian probing is used to obtain measurements of the DL channel. 
	 	\item Antenna Selection Probing\\
	 	In this scheme the BS transmits probing vectors that have zero entries except for a single entry, i.e.,
	 	\[\bphi_i^\transp = [0,\ldots,0,\sqrt{\Pbf},0,\ldots,0 ].  \]
	 	The non-zero locations are chosen uniformly at random and distinct across all probing vectors.
	 	 This is equivalent to generating a sensing matrix $\bf\Phi$ with rows that have zero entries except in a single entry which is chosen uniformly at random and has a value equal to $\sqrt{\Pbf}$. 
	 	\item Hybrid Probing\\ 
	 	In addition to the previous designs, we propose a new \textit{Hybrid} probing scheme. It has been observed that in a multi-user Massive MIMO system the user channel vectors are correlated as a result of sharing common MPCs during propagation \cite{poutanen2010significance}. This causes these channel vectors to share a portion of their support as common support. As an illustration, see Fig. \ref{supp_sets} where two support indices are common among all users (indices 3 and 8). We use this fact to propose probing vectors that provide more informative measurements. Define $\Sc_c$ to be the set of common support indices among all users, i.e., 
	 		 \begin{equation}
	 		\Sc_c = {\mbox{\larger[3]$\cap$}}_{i=1}^K \hat{\Sc}_{{\rm dl}}^{(i)}.
	 		 \end{equation}
	 		 The Hybrid probing matrix is given by
	 		 \begin{equation}\label{eq:hybrid_mat}
	 		 \bf\Phi = \left[
	 		 \begin{array}{c}
	 		 \bfF_{\Sc_c}^\herm  \\
	 		 \hline
	 		 \bfG
	 		 \end{array}
	 		 \right],
	 		 \end{equation}
	 		 
	 		 where $ \bfF_{\Sc_c}$ is a submatrix of the Fourier matrix $\bfF$, consisting of columns whose indices are in $\Sc_c$ and $\bfG$ is a matrix with Gaussian i.i.d entries. Also, the rows of the probing matrix are normalized to have an $\ell_2$ norm equal to $\sqrt{\Pbf}$. 
	 \end{itemize}
	 
	 After $T$ time slots, each user has a measurement vector of size $T$ across each sub-carrier and sends this vector back to the BS. In this paper, we will not focus on the scheme used for feeding back the vectors $\bfy^{(i)}$ and we assume that these vectors are perfectly fed back to the BS. The discussion about different feedback schemes is postponed to a future work.
	 \subsection{Channel Estimation}\label{subsec:ch_es}
	 After collecting the measurements $\yv^{(i)},~ i=1,\ldots,K$, the BS uses its estimate of the DL support of each of the users to estimate their channel. This is done by simply solving the following least squares problem,
	 \begin{equation}
	 \widehat{\check{\bfh}}_{{\rm dl}}^{(i)} = \underset{\bfx\in \bC^{M\times 1}}{\argmin} \Vert \bfy^{(i)} - {\bf\Phi} \bfF_{\suppdlhat^{(i)}} \bfx\Vert, 
	 \end{equation}
	 where $\bfF_{\suppdlhat^{(i)}}$ is a submatrix of the Fourier matrix $\bfF$ with columns whose indices are in $\suppdlhat^{(i)}$. The solution to this problem is given by
	 \begin{equation}
	 \widehat{\check{\bfh}}_{{\rm dl}}^{(i)} = \left( {\bf\Phi} \bfF_{\suppdlhat^{(i)}}  \right)^\dagger \bfy^{(i)}.
	 \end{equation}
	where $(\cdot)^\dagger$ is the Moore-Penrose pseudo-inverse. We expect that the estimated channel is very close to the original channel, since the BS uses the support information for estimation. Also, the estimation process is substantially different from the method presented in \cite{rao2014distributed}. In contrast to that work, we do not need to run any compressed sensing algorithm, because we have an explicit estimation of the channel support. 
	\subsection{Precoding}
	Let $\widehat{\bfH}_{{\rm dl}}=\bfF\left[ \widehat{\check{\bfh}}_{{\rm dl}}^{(1)}, \ldots, \widehat{\check{\bfh}}_{{\rm dl}}^{(K)}\right]$ be the matrix consisting of estimated DL channel vectors for all users. The BS uses this matrix to perform DL precoding to eliminate inter-user interference.  Here, we consider the zero-forcing (ZF) precoder, which is a matrix denoted by
	\begin{equation}
	{\bfT}_{.,j} := \frac{\Qm_{.,j} }{\Vert \Qm_{.,j} \Vert},
	\end{equation}
  where $\Qm = \left(\widehat{\bfH}_{{\rm dl}}^\herm\right)^\dagger$. The transmit signal at the BS is then given by
    \begin{equation}
    \bfx = \sqrt{\frac{P}{K}}\bfT \bfs,
    \end{equation}	
    where $\bfs \in \bC^{K\times 1}$ is the vector of unit-power user symbols $s_i$ and $P$ is the  transmit power. The received signal at user $i$ can be written as
    \begin{equation}
    r_i= \sqrt{\frac{P}{K}} \left(\bfh_{{\rm dl}}^{(i)}\right)^\herm \bfT \bfs + n_i = \sqrt{\frac{P}{K}}\left(\bfh_{{\rm dl}}^{(i)}\right)^\herm \bfT_{.,i} s_i + \sqrt{\frac{P}{K}}\underset{ j\neq i}{\sum}\left(\bfh_{{\rm dl}}^{(j)}\right)^\herm \bfT_{.,j} s_j + n_i,
    \end{equation} 
	where $n_i \sim \cg(0,1)$. The \textit{Signal to Noise plus Interference Ratio} ($\SINR$) for a user $i$ can be calculated as
	\begin{equation}\label{eq:SINR}
	\SINR_i = \frac{\frac{P}{K} \left| \left(\bfh_{{\rm dl}}^{(i)}\right)^\herm \bfT_{.,i} \right|^2}{1 + \frac{P}{K} \underset{j\neq i}{\sum}\left| \left(\bfh_{{\rm dl}}^{(j)}\right)^\herm \bfT_{.,j} \right|^2}.
	\end{equation}
	Finally, the user rate is given by
	\begin{equation}\label{eq:Rate}
	R_i = \log_2 \left( 1 + \SINR_i \right).
	\end{equation}
\section{Simulation Results}\label{sec:sim_results}
\begin{table}[t]
	\centering
	\begin{tabular}{ |l|c|c| }
		\hline
		\multicolumn{3}{ |c| }{Simulation Parameters} \\
		\hline
		Maximum Angular Range & $2\theta_{\max}$ & $\frac{2\pi}{3}$ \\ \hline
		Antenna Spacing & $d$ & $\frac{\lambda_{{\rm ul}}}{2 \sin(\theta_{\max})}$\\ \hline
		Carrier Wavelength over DL & $\lambda_{{\rm dl}}$ & $\approx \frac{\lambda_{{\rm ul}}}{1.1}$ \\ \hline
		Number of Antennas & $M$ & $256$ \\ \hline
		Number of Sketches in UL & $m$ & $64$ \\ \hline
		Number of Users & $K$ & $20$ \\ \hline
		Number of Sub-carriers & $L$ & $10$ \\ \hline
		Fourier Oversampling Factor & $q$ & $2$ \\ \hline
	\end{tabular}
	\caption{Table of simulation parameters.}
	\label{tab:table_1} 
\end{table}
In this section we provide numerical simulation results to assess the performance of our proposed algorithm empirically. We compare our algorithm with the case in which the BS has access to the full noisy channel state information (CSIT) and also with the algorithm proposed in \cite{rao2014distributed}. This algorithm uses a joint orthogonal matching pursuit (J-OMP) method to reconstruct the sparse vector of channel coefficients $\check{\hv}_{{\rm dl}}^{(i)}$ from noisy measurements $\bfy^{(i)}$.  Note that the measurements are the same for our method with Gaussian probing and the J-OMP algorithm, since this is the only channel probing scheme proposed in \cite{rao2014distributed}. With other probing methods the measurements are obviously different. We assume that for every user there are two MPCs in its communication path to the BS. This implies that the support of the scattering function consists of two subsets over the interval $\left[-\theta_{\max},\theta_{\max}\right)$. The length of the support is set to be $\left|\Xc_{\gamma}\right|\approx \frac{2\theta_{\max}}{8}$. Similar to \cite{rao2014distributed}, we assume that the users share a common MPC which results in a common channel support among all users. Nevertheless, we also investigate the effect of removing the common MPC in one of our simulations. It is important to note that since we use the COST 2100 channel model, the channel is generated according to a continuous scattering function. Therefore, when the channel is represented in the DFT basis, the corresponding vector of coefficients is not sparse in the strict sense, but rather well-approximated by a sparse vector. This is slightly different from the setting proposed in \cite{rao2014distributed} where the channel is assumed to be strictly sparse. For a fair comparison, we feed the number of dominant non-zero channel coefficients to both algorithms. We also provide the J-OMP algorithm with the size of the common support among all users, while this information is not provided to our proposed algorithm. Table \ref{tab:table_1} summarizes the main parameters used in our simulations. 	
  \begin{figure}[t]
	\centering
	\includegraphics{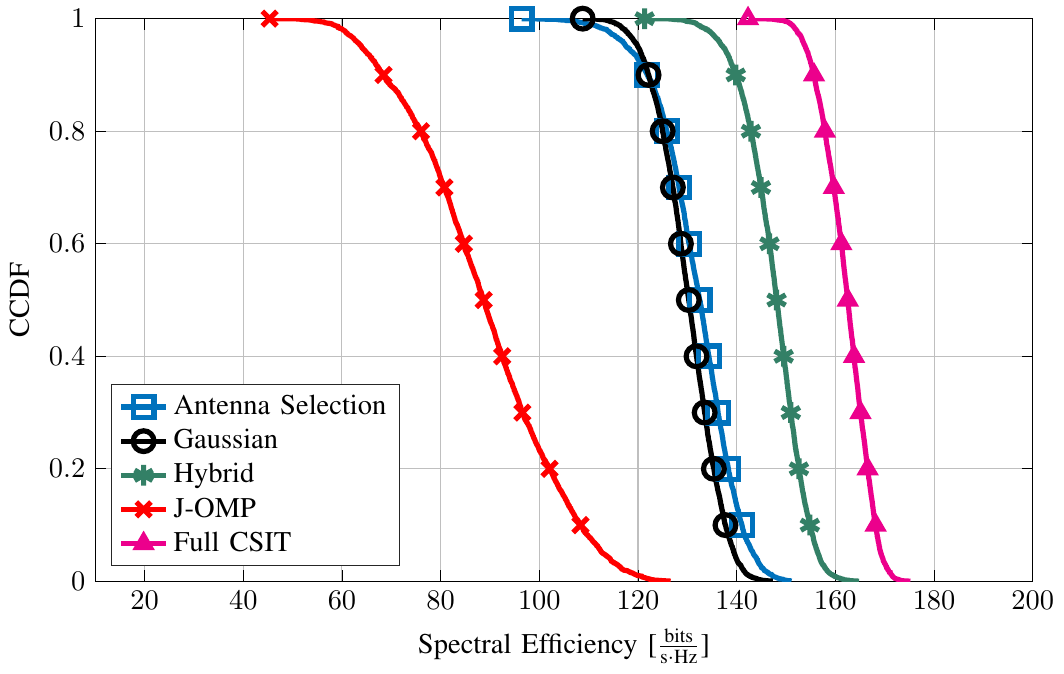}
	\caption{CCDF of spectral efficiency for our proposed algorithm with different probing schemes and the J-OMP algorithm. Here the uplink $\SNR$ is equal to 15 dB, the downlink $\SNR$ is equal to 20 dB, and the number of probing vectors is $T=80$.}
	\label{fig:Rate_1}
\end{figure}
During the simulations the users transmit their UL pilots each over $L=10$ subcarriers. The $\SNR$ for UL transmission is set to $15$ dB. Then, the BS estimates the UL angular support of each user according to the method described in \ref{sec:up_supp_es}. This gives an estimate of the angular support of the scattering function, i.e., an estimate of $\Xgam$. This information in turn determines an estimate of the angular support of each user in the DL. We use the DL support estimate both to design the Hybrid probing matrix and to recover the sparse channel coefficients from the noisy set of measurements. Channel probing is carried out via all three types of probing matrix designs as described in \ref{subsec:sensing_schemes} and we use our proposed method to estimate the channel. For the J-OMP algorithm we only use the Gaussian probing matrix, since this is the only sensing matrix considered in \cite{rao2014distributed}. After estimating the channel as described in \ref{subsec:ch_es}, we construct the ZF precoding matrix for our proposed method, the J-OMP method, and the baseline full CSIT case. The ZF precoding matrices are then used to transmit in the DL. We calculate the $\SINR$ and rate according to the formulas \eqref{eq:SINR} and \eqref{eq:Rate}. The sum rate for a single simulation can be calculated as 
\begin{equation}
\text{Sum-Rate} = \sum_{i=1}^{K} R_i,
\end{equation}
 and gives a reasonable metric to compare all the methods.
  \begin{figure}[t]
 	\centering
\includegraphics{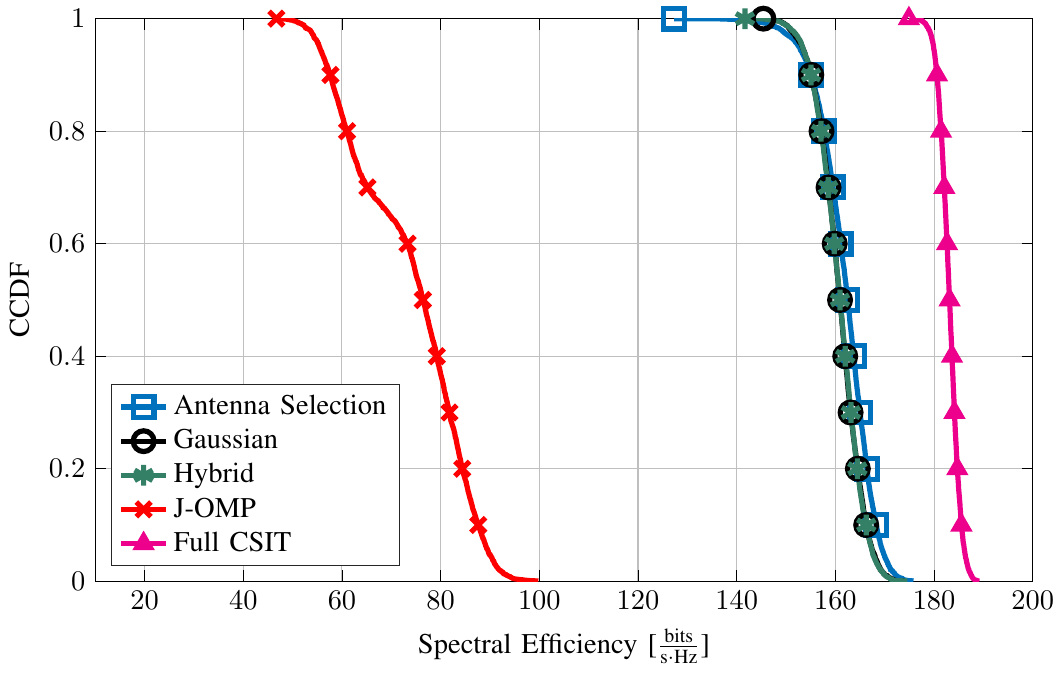}
 	\caption{CCDF of spectral efficiency for our proposed algorithm with different probing schemes and the J-OMP algorithm, without common support among the users. Here the uplink $\SNR$ is equal to 15 dB, the downlink $\SNR$ is equal to 20 dB, and the number of probing vectors is $T=80$.}
 	\label{fig:Rate_without_comm_supp}
 \end{figure}  
\subsection{Complementary CDF of the Spectral Efficiency}
Fig. \ref{fig:Rate_1} illustrates the empirical \textit{Complementary Cumulative Distribution Function} (CCDF) curve of the random spectral efficiency value  defined as the sum-rate per second per Hertz for our proposed method with all three types of probing matrices, the J-OMP method with Gaussian probing matrix and the baseline full-CSIT method. As we can see, our proposed method enjoys a considerable improvement compared with the one proposed in \cite{rao2014distributed}. The main reason is that we use the UL signals to estimate the angular support of the scattering function which results in a much better channel estimation quality. A higher rate is achieved via the Hybrid probing scheme, since the measurements corresponding to the common support coefficients are much more efficient compared with the Gaussian measurements and they introduce less noise to the recovery algorithm. \\
In addition we studied the effect of removing the common MPC on rate performance. In this experiment we generated two MPCs for each user with completely random locations and with no fixed common MPC among the users. Fig. \ref{fig:Rate_without_comm_supp} illustrates the CCDF of spectral efficiency for this experiment. As we can see, in this case, unlike our proposed method the J-OMP algorithm degrades in performance. This behavior arises because the J-OMP algorithm is based on the assumption of a common support among all users and when this assumption is violated, channel estimation has a lower quality. In addition, in the case of our proposed algorithm, the Gaussian and Hybrid probing methods have more or less the same performance, since without a common support among users, the proposed design of a Hybrid probing matrix in \eqref{eq:hybrid_mat} coincides with that of a Gaussian probing matrix.
Generally, when a fixed common MPC is not present, the rate performance for methods other than the J-OMP method increases slightly. The reason is that in this case the channel vectors are less correlated compared with the case where a common MPC is present. As a result, with ZF precoding, the received power to each user is greater than when we have a common MPC. This will improve $\SINR$ which in turn increases the rate.
\begin{figure}[t]
	\centering
	\includegraphics{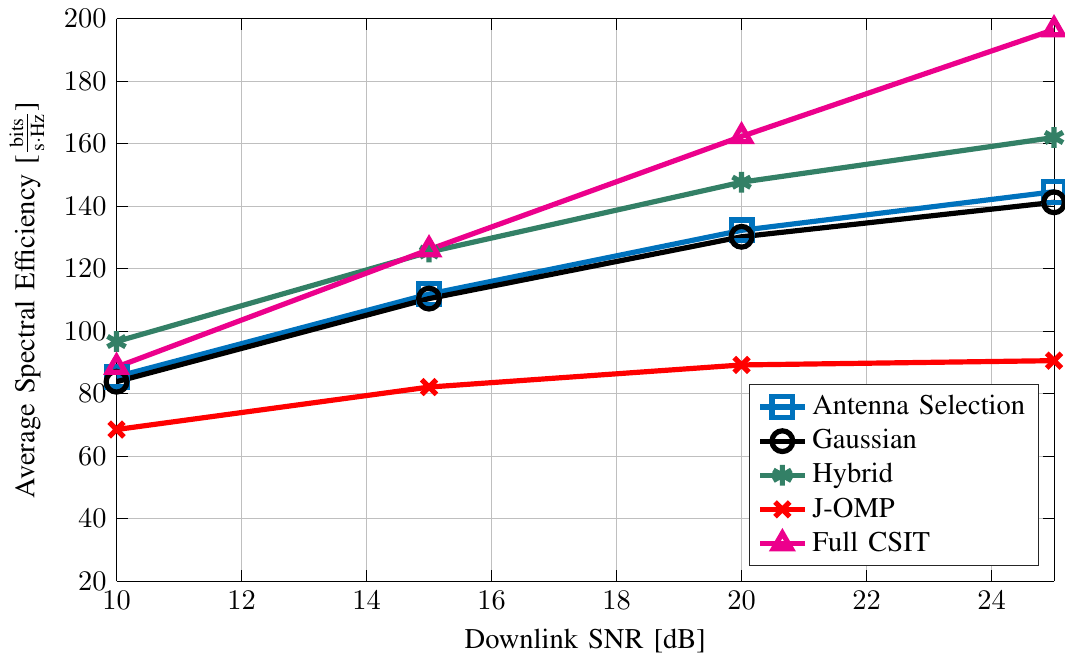}
	\caption{Average spectral efficiency for our proposed algorithm with different probing schemes and the J-OMP algorithm vs downlink $\SNR$. Here the uplink $\SNR$ is equal to 15 dB, and the number of probing vectors is $T=80$.}
	\label{fig:Rate_vs_SNR}
\end{figure}
\subsection{Effect of Downlink SNR}
Fig. \ref{fig:Rate_vs_SNR} compares the performance of different schemes as a function of the downlink $\SNR$. This $\SNR$ value effects the recovery because with noisier measurements the performance degrades. Here we average the rate value over 2000 Monte-Carlo simulations for each $\SNR$ value. One interesting observation is that in low $\SNR$ values, our proposed Hybrid probing method achieves a higher spectral efficiency even compared to the full CSIT scenario. The reason is that we use the additional support information to estimate the channel vector, which reduces the noise effect by limiting the signal subspace to basis vectors whose indices are in the support set. With higher $\SNR$, this comparison changes. As we described before, the channel is not sparse in a strict sense and as a result there are non-zero coefficients outside the support set. In other words, the signal power \textit{leaks} out of the support set, which consists of the indices of dominant coefficients. This leakage effects performance in high $\SNR$ regimes and dominates the noise effect. Therefore, we see that the spectral efficiency in estimation methods saturates in high $\SNR$ values, whereas for the full CSIT case, it increases linearly.

\subsection{Effect of the Number of Measurements}
\begin{figure}[t]
	\centering
\includegraphics{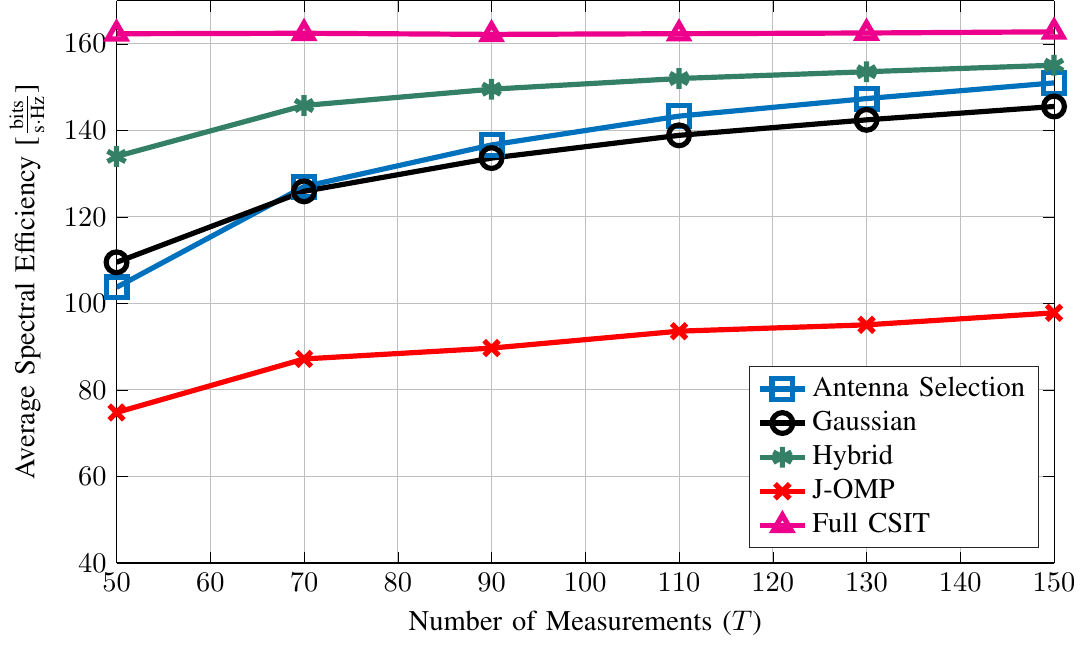}
	\caption{Average spectral efficiency for our proposed algorithm with different probing schemes and the J-OMP algorithm vs the number of $T$. Here the uplink $\SNR$ is equal to $15$ dB, and the downlink $\SNR$ is equal to $20$ dB.}
	\label{fig:Rate_vs_T}
\end{figure}

Fig. \ref{fig:Rate_vs_T} compares the results as a function of the number of channel probings (or measurements), which indicates the feedback overhead. It can be seen that there exists a substantial gap between the performance of our proposed algorithm in terms of rate with that of the J-OMP algorithm.

\section{Conclusion}
We proposed a novel downlink channel probing and estimation method in a FDD massive MIMO scenario. This method is based on the observation that the spatial scattering function is invariant with respect to the carrier frequency. We used the uplink channel vectors to estimate the support of this scattering function which in turn gives an estimate of the support of the downlink channel vectors. This information helps design a new channel probing scheme, reduces the number of necessary channel probings, reduces the feedback overhead and therefore makes implementing the FDD massive MIMO system a feasible idea. Our empirical results show that the proposed method is superior to the existing compressed sensing techniques in both the channel estimation quality and reduction in feedback overhead.

\balance
\bibliographystyle{IEEEtran}
{\small
\bibliography{references}}

\end{document}